\documentclass[10pt,a4paper,onecolumn]{article}
\usepackage{marginnote}
\usepackage{graphicx}
\usepackage{xcolor}
\usepackage{authblk,etoolbox}
\usepackage{titlesec}
\usepackage{calc}
\usepackage{tikz}
\usepackage{hyperref}
\hypersetup{colorlinks,breaklinks,
            urlcolor=[rgb]{0.0, 0.5, 1.0},
            linkcolor=[rgb]{0.0, 0.5, 1.0}}
\usepackage{caption}
\usepackage{tcolorbox}
\usepackage{amssymb,amsmath}
\usepackage{ifxetex,ifluatex}
\usepackage{seqsplit}
\usepackage{xstring}

\usepackage{float}
\let\origfigure\figure
\let\endorigfigure\endfigure
\renewenvironment{figure}[1][2] {
    \expandafter\origfigure\expandafter[H]
} {
    \endorigfigure
}

\usepackage{fixltx2e} 
\usepackage[
  backend=biber,
]{biblatex}
\bibliography{paper.bib}

\let\textttOrig=\texttt
\def\texttt#1{\expandafter\textttOrig{\seqsplit{#1}}}
\renewcommand{\seqinsert}{\ifmmode
  \allowbreak
  \else\penalty6000\hspace{0pt plus 0.02em}\fi}

\makeatletter
\let\href@Orig=\href
\def\href@Urllike#1#2{\href@Orig{#1}{\begingroup
    \def\Url@String{#2}\Url@FormatString
    \endgroup}}
\def\href@Notdoi#1#2{\def\tempa{#1}\def\tempb{#2}%
  \ifx\tempa\tempb\relax\href@Urllike{#1}{#2}\else
  \href@Orig{#1}{#2}\fi}
\def\href#1#2{%
  \IfBeginWith{#1}{https://doi.org}%
  {\href@Urllike{#1}{#2}}{\href@Notdoi{#1}{#2}}}
\makeatother

\usepackage[top=3.5cm, bottom=3cm, right=1.5cm, left=1.0cm,
            headheight=2.2cm, reversemp, includemp, marginparwidth=4.5cm]{geometry}

\titleformat{\section}
  {\normalfont\sffamily\Large\bfseries}
  {}{0pt}{}
\titleformat{\subsection}
  {\normalfont\sffamily\large\bfseries}
  {}{0pt}{}
\titleformat{\subsubsection}
  {\normalfont\sffamily\bfseries}
  {}{0pt}{}
\titleformat*{\paragraph}
  {\sffamily\normalsize}

\usepackage{fancyhdr}
\makeatletter
\let\ps@plain\ps@fancy
\fancyheadoffset[L]{4.5cm}
\fancyfootoffset[L]{4.5cm}

\definecolor{linky}{rgb}{0.0, 0.5, 1.0}

\newtcolorbox{repobox}
   {colback=red, colframe=red!75!black,
     boxrule=0.5pt, arc=2pt, left=6pt, right=6pt, top=3pt, bottom=3pt}

\newcommand{\ExternalLink}{%
   \tikz[x=1.2ex, y=1.2ex, baseline=-0.05ex]{%
       \begin{scope}[x=1ex, y=1ex]
           \clip (-0.1,-0.1)
               --++ (-0, 1.2)
               --++ (0.6, 0)
               --++ (0, -0.6)
               --++ (0.6, 0)
               --++ (0, -1);
           \path[draw,
               line width = 0.5,
               rounded corners=0.5]
               (0,0) rectangle (1,1);
       \end{scope}
       \path[draw, line width = 0.5] (0.5, 0.5)
           -- (1, 1);
       \path[draw, line width = 0.5] (0.6, 1)
           -- (1, 1) -- (1, 0.6);
       }
   }

\patchcmd{\@maketitle}{center}{flushleft}{}{}
\patchcmd{\@maketitle}{center}{flushleft}{}{}
\patchcmd{\@maketitle}{\LARGE}{\LARGE\sffamily}{}{}
\def\maketitle{{%
  
  \AB@maketitle}}
\makeatletter
\renewcommand\AB@affilsepx{ \protect\Affilfont}
\renewcommand\AB@affilnote[1]{{\bfseries #1}\hspace{3pt}}
\renewcommand{\affil}[2][]%
   {\newaffiltrue\let\AB@blk@and\AB@pand
      \if\relax#1\relax\def\AB@note{\AB@thenote}\else\def\AB@note{#1}%
        \setcounter{Maxaffil}{0}\fi
        \begingroup
        \let\href=\href@Orig
        \let\texttt=\textttOrig
        \let\protect\@unexpandable@protect
        \def\thanks{\protect\thanks}\def\footnote{\protect\footnote}%
        \@temptokena=\expandafter{\AB@authors}%
        {\def\\{\protect\\\protect\Affilfont}\xdef\AB@temp{#2}}%
         \xdef\AB@authors{\the\@temptokena\AB@las\AB@au@str
         \protect\\[\affilsep]\protect\Affilfont\AB@temp}%
         \gdef\AB@las{}\gdef\AB@au@str{}%
        {\def\\{, \ignorespaces}\xdef\AB@temp{#2}}%
        \@temptokena=\expandafter{\AB@affillist}%
        \xdef\AB@affillist{\the\@temptokena \AB@affilsep
          \AB@affilnote{\AB@note}\protect\Affilfont\AB@temp}%
      \endgroup
       \let\AB@affilsep\AB@affilsepx
}
\makeatother

\renewcommand\Affilfont{\sffamily\small\mdseries}
\ifnum 0\ifxetex 1\fi\ifluatex 1\fi=0 
  \usepackage[T1]{fontenc}
  \usepackage[utf8]{inputenc}

\else 
  \ifxetex
    \usepackage{mathspec}
  \else
    \usepackage{fontspec}
  \fi
  \defaultfontfeatures{Ligatures=TeX,Scale=MatchLowercase}

\fi
\IfFileExists{upquote.sty}{\usepackage{upquote}}{}
\IfFileExists{microtype.sty}{%
\usepackage{microtype}
\UseMicrotypeSet[protrusion]{basicmath} 
}{}

\usepackage{hyperref}
\hypersetup{unicode=true,
            pdftitle={deeplenstronomy: A dataset simulation package for strong gravitational lensing},
            pdfborder={0 0 0},
            breaklinks=true}
\let\addcontentslineOrig=\addcontentsline
\def\addcontentsline#1#2#3{\bgroup
  \let\texttt=\textttOrig\addcontentslineOrig{#1}{#2}{#3}\egroup}
\let\markbothOrig\markboth
\def\markboth#1#2{\bgroup
  \let\texttt=\textttOrig\markbothOrig{#1}{#2}\egroup}
\let\markrightOrig\markright
\def\markright#1{\bgroup
  \let\texttt=\textttOrig\markrightOrig{#1}\egroup}

\usepackage{graphicx,grffile}
\makeatletter
\def\maxwidth{\ifdim\Gin@nat@width>\linewidth\linewidth\else\Gin@nat@width\fi}
\def\maxheight{\ifdim\Gin@nat@height>\textheight\textheight\else\Gin@nat@height\fi}
\makeatother
\setkeys{Gin}{width=\maxwidth,height=\maxheight,keepaspectratio}
\IfFileExists{parskip.sty}{%
\usepackage{parskip}
}{
\setlength{\parindent}{0pt}
\setlength{\parskip}{6pt plus 2pt minus 1pt}
}
\ifx\paragraph\undefined\else
\let\oldparagraph\paragraph
\renewcommand{\paragraph}[1]{\oldparagraph{#1}\mbox{}}
\fi
\ifx\subparagraph\undefined\else
\let\oldsubparagraph\subparagraph
\renewcommand{\subparagraph}[1]{\oldsubparagraph{#1}\mbox{}}
\fi

\title{deeplenstronomy: A dataset simulation package for strong gravitational
lensing}

        \author[1, 2]{Robert Morgan\footnote{Corresponding author}}
          \author[3, 4]{Brian Nord}
          \author[5]{Simon Birrer}
          \author[6]{Joshua Yao-Yu Lin}
          \author[4]{Jason Poh}
    
      \affil[1]{University of Wisconsin-Madison}
      \affil[2]{Legacy Survey of Space and Time Data Science Fellowship Program}
      \affil[3]{Fermi National Accelerator Laboratory}
      \affil[4]{University of Chicago}
      \affil[5]{Stanford University}
      \affil[6]{University of Illinois Urbana-Champaign}
  \date{\vspace{-5ex}}

\begin{document}
\maketitle

\marginpar{
  \sffamily\small

  {\bfseries DOI:} \href{https://doi.org/10.21105/joss.02854}{\color{linky}{10.21105/joss.02854}}

  \vspace{2mm}

  {\bfseries Software}
  \begin{itemize}
    \setlength\itemsep{0em}
    \item \href{https://github.com/openjournals/joss-reviews/issues/2854}{\color{linky}{Review}} \ExternalLink
    \item \href{https://github.com/deepskies/deeplenstronomy}{\color{linky}{Repository}} \ExternalLink
    \item \href{https://doi.org/10.5281/zenodo.4479712}{\color{linky}{Archive}} \ExternalLink
  \end{itemize}

  \vspace{2mm}

  {\bfseries Submitted:} 10 November 2020\\
  {\bfseries Published:} 04 February 2021

  \vspace{2mm}
  {\bfseries License}\\
  Authors of papers retain copyright and release the work under a Creative Commons Attribution 4.0 International License (\href{https://creativecommons.org/licenses/by/4.0/}{\color{linky}{CC BY 4.0}}).
}

\section{Background}\label{background}

Astronomical observations and statistical modeling permit the
high-fidelity analysis of strong gravitational lensing (SL) systems,
which display an astronomical phenomenon in which light from a distant
object is deflected by the gravitational field of another object along
its path to the observer. These systems are of great scientific interest
because they provide information about multiple astrophysical and
cosmological phenomena, including the nature of dark matter, the
expansion rate of the Universe, and characteristics of galaxy
populations. They also serve as standing tests of the theory of General
Relativity and modified theories of gravity.

Traditional searches for SL systems have involved time- and
effort-intensive visual or manual inspection of images by humans to
identify characteristic features --- like arcs, particular color
combinations, and object orientations. However, a comprehensive search
using the traditional approach is prohibitively expensive for large
numbers of images, like those in cosmological surveys --- e.g., the
Sloan Digital Sky Survey (\hyperlink{ref-sdss}{York et al. 2000}), the Dark Energy Survey
(\hyperlink{ref-des}{Abbott et al. 2018}), and the Legacy Survey of Space and Time (LSST)
(\hyperlink{ref-lsst}{Ivezić et al. 2019}). To automate the SL detection process, techniques
based on machine learning (ML) are beginning to overtake traditional
approaches for scanning astronomical images. In particular, deep
learning techniques have been the focus, but they require large sets of
labeled images to train these models. Because of the relatively low
number of observed SL systems, simulated datasets of images are often
needed. Thus, the composition and production of these simulated datasets
have become integral parts of the SL detection process.

One of the premier tools for simulating and analyzing SL systems,
\texttt{lenstronomy} (\hyperlink{ref-lenstronomy}{Birrer and Amara 2018}), works by the user
specifying the properties of the physical systems, as well as how they
are observed (e.g., telescope and camera) through a
\texttt{python}-based application programming interface (API) to
generate a single image. Generating populations of SL systems that are
fit for neural network training requires additional infrastructure.

\section{Statement of need}\label{statement-of-need}

Due to the inherent dependence of the performance of ML approaches on
their training data, the deep learning approach to SL detection is in
tension with scientific reproducibility without a clear prescription for
the simulation of the training data. There is a critical need for a tool
that simulates full datasets in an efficient and reproducible manner,
while enabling the use of all the features of the \texttt{lenstronomy}
simulation API. Additionally, this tool should simplify user interaction
with \texttt{lenstronomy} and organize the simulations and associated
metadata into convenient data structures for deep learning problems.
Multiple packages have been developed to generate realistic training
data by wrapping around \texttt{lenstronomy}: \texttt{baobab} (\hyperlink{ref-baobab}{Park
2021}) generates training sets for lens modeling and hierarchical
inference and the LSST Dark Energy Science Collaboration's
\texttt{SLSprinkler} (\hyperlink{ref-lsstdescsl}{Kalmbach et al. 2020}) adds strongly lensed
variable objects into catalogs and images. Nonetheless, the need for a
simple, general tool capable of efficiently simulating any astronomical
system in a reproducible manner while giving the user complete freedom
to set the properties of objects remains.

\section{Summary}\label{summary}

\texttt{deeplenstronomy} generates SL datasets by organizing and
expediting user interaction with \texttt{lenstronomy}. The user creates
a single yaml-style configuration file that describes the aspects of the
dataset: number of images, properties of the telescope and camera,
cosmological parameters, observing conditions, properties of the
physical objects, and geometry of the SL systems.
\texttt{deeplenstronomy} parses the configuration file and generates the
dataset, producing both the images and the parameters that led to the
production of each image as outputs. The configuration files can easily
be shared, enabling users to easily reproduce each other's training
datasets.

\begin{figure}
\centering
\includegraphics{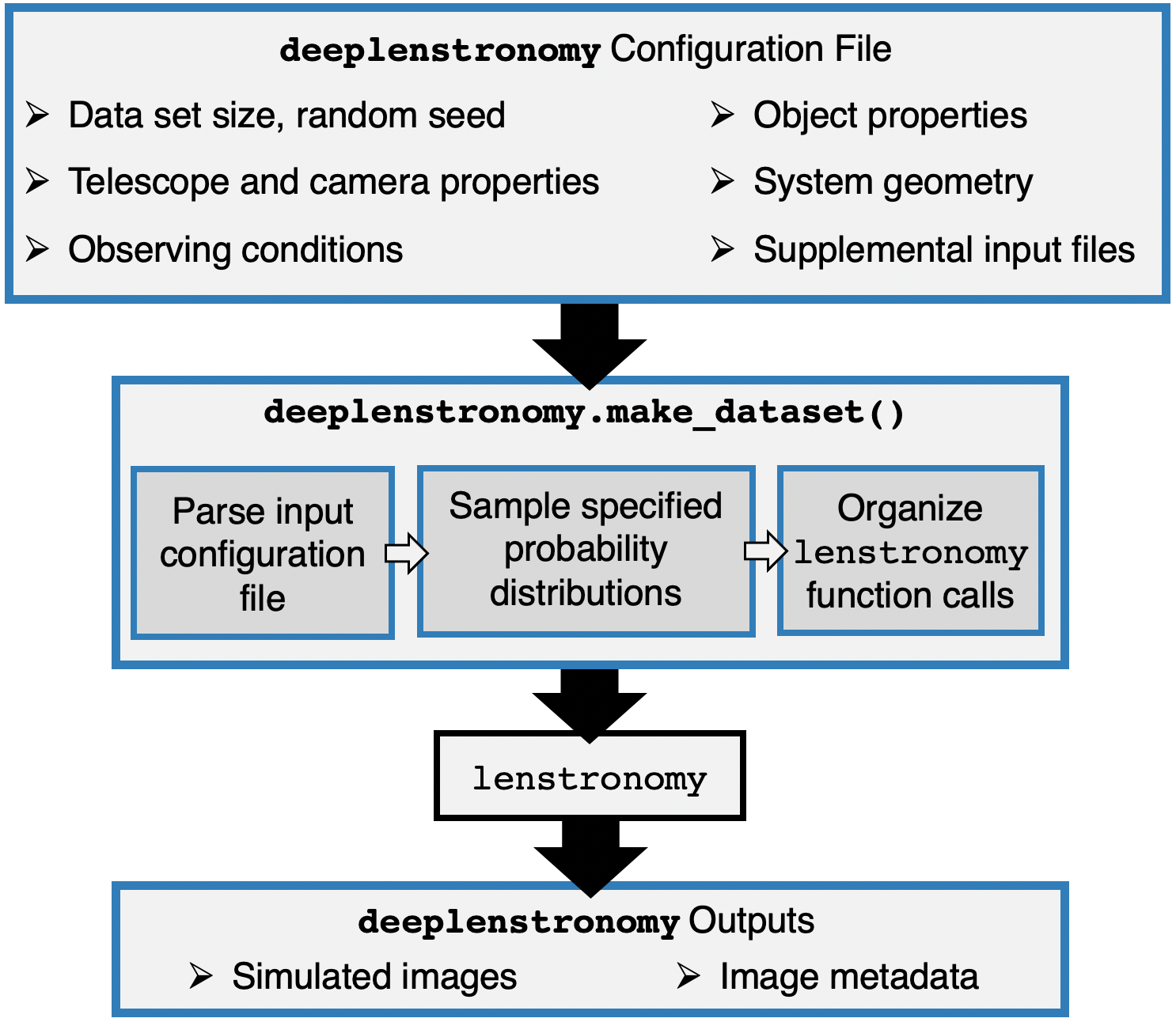}
\caption{The \texttt{deeplenstronomy} process. Dataset properties,
camera and telescope properties, observing conditions, object properties
(e.g., \texttt{lenstronomy} light and mass profiles, point sources, and
temporal behavior), the geometry of the SL systems, and optional
supplemental input files (e.g., probability distributions, covariance
matrices, and image backgrounds) are specified in the main configuration
file. \texttt{deeplenstronomy} then interprets the configuration file,
calls \texttt{lenstronomy} simulation functionalities, and organizes the
resulting images and metadata.\label{fig:flowchart}}
\end{figure}

The premier objective of \texttt{deeplenstronomy} is to help astronomers
make their training datasets as realistic as possible. To that end,
\texttt{deeplenstronomy} contains built-in features for the following
functionalities: use any stellar light profile or mass profile in
\texttt{lenstronomy}; simulate a variety of astronomical systems such as
single galaxies, foreground stars, galaxy clusters, supernovae, and
kilonovae, as well as any combination of those systems; fully control
the placements of objects in the simulations; use observing conditions
of real astronomical surveys; draw any parameter from any probability
distribution; introduce any correlation; and incorporate real images
into the simulation. Furthermore, \texttt{deeplenstronomy} facilitates
realistic time-domain studies by providing access to public spectral
energy distributions of observed supernovae and kilonovae and
incorporating the transient objects into time series of simulated
images. Finally, \texttt{deeplenstronomy} provides data visualization
functions to enable users to inspect their simulation outputs. These
features and the path from configuration file to full data set are shown
in \autoref{fig:flowchart}.

\texttt{deeplenstronomy} makes use of multiple open-source software
packages: \texttt{lenstronomy} is used for all gravitational lensing
calculations and image simulation; \texttt{numpy} (\hyperlink{ref-numpy}{Harris et al. 2020})
\texttt{Array}s are used internally to store image data and perform
vectorized calculations; \texttt{pandas} (\hyperlink{ref-pandas}{McKinney et al. 2010})
\texttt{DataFrame}s are utilized for storing simulation metadata and
file reading and writing; \texttt{scipy} (\hyperlink{ref-scipy}{Jones et al. 2001}) is used for
integration and interpolation; \texttt{matplotlib} (\hyperlink{ref-matplotlib}{Hunter 2007})
functions are used for image visualization; \texttt{astropy} (\hyperlink{ref-astropy}{Astropy
Collaboration et al. 2013}) is used for cosmological calculations and
color image production; \texttt{h5py} (\hyperlink{ref-h5py}{Collette 2014}) is utilized for
saving images; and \texttt{PyYAML} (\hyperlink{ref-pyyaml}{Simonov and Net 2006}) is used to
manage the configuration file. While not used directly, some
\texttt{python-benedict} (\hyperlink{ref-benedict}{Caccamo 2018}) functionalities helped to create
\texttt{deeplenstronomy}'s data structures and internal search
algorithms.

\texttt{deeplenstronomy} is packaged and disseminated via
\href{https://pypi.org/project/deeplenstronomy/}{PyPI}. Documentation
and example notebooks are available on the
\href{https://deepskies.github.io/deeplenstronomy/}{\texttt{deeplenstronomy}
website}. Any bugs or feature requests can be opened as issues in the
\href{https://github.com/deepskies/deeplenstronomy/issues}{GitHub
repository} (\hyperlink{ref-deeplenstronomy}{Morgan 2020}).

\section{Acknowledgements}\label{acknowledgements}

R. Morgan thanks the LSSTC Data Science Fellowship Program, which is
funded by LSSTC, NSF Cybertraining Grant \#1829740, the Brinson
Foundation, and the Moore Foundation; his participation in the program
has benefited this work. R. Morgan also thanks the Universities Research
Association Fermilab Visiting Scholar Program for funding his work on
this project.

We acknowledge the Deep Skies Lab as a community of multi-domain experts
and collaborators who've facilitated an environment of open discussion,
idea-generation, and collaboration. This community was important for the
development of this project. We acknowledge contributions from Joao
Caldeira during the early stages of this project.

Work supported by the Fermi National Accelerator Laboratory, managed and
operated by Fermi Research Alliance, LLC under Contract No.
DE-AC02-07CH11359 with the U.S. Department of Energy. The U.S.
Government retains and the publisher, by accepting the article for
publication, acknowledges that the U.S. Government retains a
non-exclusive, paid-up, irrevocable, world-wide license to publish or
reproduce the published form of this manuscript, or allow others to do
so, for U.S. Government purposes.

\section*{References}\label{references}
\addcontentsline{toc}{section}{References}

\hypertarget{refs}{}
\hypertarget{ref-des}{}
Abbott, T. M. C., F. B. Abdalla, A. Alarcon, J. Aleksić, S. Allam, S.
Allen, A. Amara, et al. 2018. ``Dark Energy Survey year 1 results:
Cosmological constraints from galaxy clustering and weak lensing'' 98
(4): 043526.
doi:\href{https://doi.org/10.1103/PhysRevD.98.043526}{10.1103/PhysRevD.98.043526}.

\hypertarget{ref-astropy}{}
Astropy Collaboration, T. P. Robitaille, E. J. Tollerud, P. Greenfield,
M. Droettboom, E. Bray, T. Aldcroft, et al. 2013. ``Astropy: A community
Python package for astronomy'' 558 (October): A33.
doi:\href{https://doi.org/10.1051/0004-6361/201322068}{10.1051/0004-6361/201322068}.

\hypertarget{ref-lenstronomy}{}
Birrer, Simon, and Adam Amara. 2018. ``Lenstronomy: Multi-Purpose
Gravitational Lens Modelling Software Package.'' \emph{Physics of the
Dark Universe} 22: 189--201.
doi:\href{https://doi.org/10.1016/j.dark.2018.11.002}{10.1016/j.dark.2018.11.002}.

\hypertarget{ref-benedict}{}
Caccamo, Fabio. 2018. ``Python-Benedict.''
\url{https://github.com/fabiocaccamo/python-benedict}.

\hypertarget{ref-h5py}{}
Collette, Andrew. 2014. \emph{Python and Hdf5}. O'Reilly.

\hypertarget{ref-numpy}{}
Harris, Charles R., K. Jarrod Millman, Stéfan J van der Walt, Ralf
Gommers, Pauli Virtanen, David Cournapeau, Eric Wieser, et al. 2020.
``Array Programming with NumPy.'' \emph{Nature} 585: 357--62.
doi:\href{https://doi.org/10.1038/s41586-020-2649-2}{10.1038/s41586-020-2649-2}.

\hypertarget{ref-matplotlib}{}
Hunter, John D. 2007. ``Matplotlib: A 2d Graphics Environment.''
\emph{Computing in Science and Engineering} 9 (3): 90--95.
doi:\href{https://doi.org/10.1109/MCSE.2007.55}{10.1109/MCSE.2007.55}.

\hypertarget{ref-lsst}{}
Ivezić, Željko, Steven M. Kahn, J. Anthony Tyson, Bob Abel, Emily
Acosta, Robyn Allsman, David Alonso, et al. 2019. ``LSST: From Science
Drivers to Reference Design and Anticipated Data Products'' 873 (2):
111.
doi:\href{https://doi.org/10.3847/1538-4357/ab042c}{10.3847/1538-4357/ab042c}.

\hypertarget{ref-scipy}{}
Jones, Eric, Travis Oliphant, Pearu Peterson, and others. 2001. ``SciPy:
Open Source Scientific Tools for Python.'' \url{http://www.scipy.org/}.

\hypertarget{ref-lsstdescsl}{}
Kalmbach, J. Bryce, Ji Won Park, Matthew Wiesner, and James Chiang.
2020. \emph{LSSTDESC/SLSprinkler: LSST DESC Strong Lensing Sprinkler for
Simulations} (version v1.1.0). Zenodo.
doi:\href{https://doi.org/10.5281/zenodo.4480392}{10.5281/zenodo.4480392}.

\hypertarget{ref-pandas}{}
McKinney, Wes, and others. 2010. ``Data Structures for Statistical
Computing in Python.'' In \emph{Proceedings of the 9th Python in Science
Conference}, 445:51--56. Austin, TX.
doi:\href{https://doi.org/10.25080/majora-92bf1922-00a}{10.25080/majora-92bf1922-00a}.

\hypertarget{ref-deeplenstronomy}{}
Morgan, Robert. 2020. ``Deeplenstronomy.''
\url{https://github.com/deepskies/deeplenstronomy}.

\hypertarget{ref-baobab}{}
Park, Ji Won. 2021. \emph{Jiwoncpark/Baobab: V0.1.2} (version v0.1.2).
Zenodo.
doi:\href{https://doi.org/10.5281/zenodo.4476822}{10.5281/zenodo.4476822}.

\hypertarget{ref-pyyaml}{}
Simonov, K., and I. Net. 2006. ``PyYaml.''
\url{https://github.com/yaml/pyyaml}.

\hypertarget{ref-sdss}{}
York, Donald G., J. Adelman, Jr. Anderson John E., Scott F. Anderson,
James Annis, Neta A. Bahcall, J. A. Bakken, et al. 2000. ``The Sloan
Digital Sky Survey: Technical Summary'' 120 (3): 1579--87.
doi:\href{https://doi.org/10.1086/301513}{10.1086/301513}.

\end{document}